\documentclass[aps,pra,twocolumn,superscriptaddress,amsmath,amssymb]{revtex4-1}

\usepackage{graphicx}
\usepackage{xspace}
\usepackage{xcolor}

\newcommand{\Fref}[1]{Figure~\ref{#1}}
\newcommand{\fref}[1]{Fig.~\ref{#1}}

\newcommand{\hc}{$H_{c1}$\xspace}

\begin{document}

\title{Quantum phase transition inside the superconducting dome of\\ Ba(Fe$_{1-x}$Co$_x$)$_2$As$_2$ from diamond-based optical magnetometry}

\author{K.~R.~Joshi}
\affiliation{Ames Laboratory, Ames, IA 50011}
\affiliation{Department of Physics $\&$ Astronomy, Iowa State University, Ames, IA 50011}

\author{N.~M.~Nusran}
\affiliation{Ames Laboratory, Ames, IA 50011}

\author{M.~A.~Tanatar}
\affiliation{Ames Laboratory, Ames, IA 50011}
\affiliation{Department of Physics $\&$ Astronomy, Iowa State University, Ames, IA 50011}

\author{K. Cho}
\affiliation{Ames Laboratory, Ames, IA 50011}

\author{S. L. Bud'ko}
\affiliation{Ames Laboratory, Ames, IA 50011}
\affiliation{Department of Physics $\&$ Astronomy, Iowa State University, Ames, IA 50011}

\author{P. C. Canfield}
\affiliation{Ames Laboratory, Ames, IA 50011}
\affiliation{Department of Physics $\&$ Astronomy, Iowa State University, Ames, IA 50011}

\author{R.~M.~Fernandes}
\affiliation{School of Physics \& Astronomy, University of Minnesota, Minneapolis, MN 55455}

\author{A.~Levchenko}
\affiliation{Physics Department, University of Wisconsin-Madison, Madison, WI  53706}

\author{R. Prozorov}
\email[Corresponding author: ]{prozorov@ameslab.gov}
\affiliation{Ames Laboratory, Ames, IA 50011}
\affiliation{Department of Physics $\&$ Astronomy, Iowa State University, Ames, IA 50011}

\date{27 August 2019}

\begin{abstract}
 Unconventional superconductivity often emerges in close proximity to a magnetic instability. Upon suppressing the magnetic transition down to zero temperature by tuning the carrier concentration, pressure, or disorder, the superconducting transition temperature $T_c$ acquires its maximum value. A major challenge is the elucidation of the relationship between the superconducting phase and the strong quantum fluctuations expected near a quantum phase transition (QPT) that is either second order (i.e. a quantum critical point) or weakly first order. While unusual normal state properties, such as non-Fermi liquid behavior of the resistivity, are commonly associated with strong quantum fluctuations, evidence for its presence inside the superconducting dome are much scarcer. In this paper, we use sensitive and minimally invasive optical magnetometry based on NV-centers in diamond to probe the doping evolution of the $T=0$ penetration depth in the electron-doped iron-based superconductor Ba(Fe$_{1-x}$Co$_x$)$_2$As$_2$. A non-monotonic evolution with a pronounced peak in the vicinity of the putative magnetic QPT is found. This behavior is reminiscent to that previously seen in isovalently-substituted BaFe$_2$(As$_{1-x}$P$_x$)$_2$ compounds, despite the notable differences between these two systems. Whereas the latter is a very clean system that displays nodal superconductivity and a single simultaneous first-order nematic-magnetic transition, the former is a charge-doped and significantly dirtier system with fully gapped superconductivity and split second-order nematic and magnetic transitions. Thus, our observation of a sharp peak in $\lambda (x) $ near optimal doping, combined with the theoretical result that a QPT alone does not mandate the appearance of such peak, unveils a puzzling and seemingly universal manifestation of magnetic quantum fluctuations in iron-based superconductors and unusually robust quantum phase transition under the dome of superconductivity.

\end{abstract}

\maketitle

\section*{Introduction}

The unconventional superconducting (SC) state of heavy fermions \cite{Mathur98, Park08, Custers2003Nature_Heavy_Fermion_QCP, Lohneysen2007RMP_QCP_HeavyFermions, KnafoFlouquet2009NatPhys_HeavyFermion_QCP, StockertSteglich2011ARCMP_QCP_HeavyFermion, YangPines2014PNAS_HeavyFermion_QCP_scaling, Jung2018NatComm_QCP_CeRhIn5, Gegenwart08}, cuprates \cite{Sachdev2003RevModPhys_Cuprates_QCP, Broun08, Vojta2009AdvInPhys_cuprates_review, Armitage2010RevModPhys_cuprates_review, Sachdev2010PhysStatusSolidi_QCP_Cuprates, Taillefer2010ARCMP_Cuprate_and_others}, organics \cite{Dressel2011JPCM_Organic_SC_QCP}, and iron-based materials \cite{Paglione08, Canfield10, Shibauchi2014ARCMP_QCP_BaP122, AnalytisFisher2014NatPhy_QCP_BaP122, PutzkeCarrington2014NatComm_BaP122_QCP_in_Hc2,Abrahams2011JPCM_review, KuoFisher2016Science_Nematic_QCP_FeSCs} is located close to the region of the phase diagram where long-range antiferromagnetism disappears, sometimes even overlapping with it. Elucidating the interplay between the putative $T=0$ antiferromagnetic (AFM) transition -- called a quantum phase transition (QPT) -- and superconductivity remains one of the main challenges in the field, particularly near a second-order QPT (also known as a quantum critical point, QCP) or a weakly first-order QPT (e.g. a fluctuation-driven first-order transition). In these cases, quantum AFM fluctuations are expected to be strong and reach quite high temperatures. As such, they are believed to manifest themselves in the non-Fermi liquid behavior of the metallic normal state and also to promote the superconducting instability by providing the glue that binds the Cooper pairs together \cite{Mazin10, Chubukov12, Scalpino12}. Throughout the manuscript, we use QPT to refer to either a second-order or a weakly first-order transition. Although both are associated with strong fluctuations, the latter only diverge in the case of a QCP.

Uncovering the presence of such a QPT is important for understanding the mechanism of unconventional superconductivity \cite{Broun08,Gegenwart08}. However, most of the probes for strong fluctuations arising from a QPT focus on its possible normal-state manifestations, such as a divergence of the effective electronic mass $m^*$ or non-$T^2$ behavior of the resistivity \cite{Shibauchi2014ARCMP_QCP_BaP122}. In contrast, only a few experimental approaches exist which may probe the impact of a QPT directly inside the SC dome. The London penetration depth, $\lambda$, is believed to be one of them. The qualitative argument is that, at zero temperature, both in the clean and dirty limits, $\lambda^{-2}(T=0) \propto 1/m^*$. It should be reminded that for a Galilean-invariant single-component superfluid, the interaction effects do not renormalize the effective mass, hence bare band mass enters the expression for $\lambda(0)$ \cite{Leggett-1965}. However, in a general case of a multi-band system, partial non-cancellation between self-energy and vertex corrections leads to the interaction-dependent renormalization of $\lambda(0)$, which makes it a sensitive probe of quantum critical fluctuations.
Of course, experimentally, the penetration depth is measured at finite temperatures, but since the variation of $\lambda(T)$ is very small up to a significant fraction of $T/T_c$, one can use $\lambda(T \ll T_c)$ instead of $\lambda(0)$.

Taking the proportionality between $\lambda^{2}$ and $m^*$ at a face value, one would expect that the presence of a second-order or weakly first-order QPT inside the SC dome would be manifested as a sharp peak in the $T=0$ penetration depth. Such a sharp peak in $\lambda$ was indeed observed under the dome of superconductivity in isovalently-substituted BaFe$_2$(As$_{1-x}$P$_x$)$_2$ (P-Ba122) \cite{Hashimoto12, Lamhot15}, and corroborated by anomalous behaviors of the critical magnetic fields $H_{c2}$ and $H_{c1}$ \cite{PutzkeCarrington2014NatComm_BaP122_QCP_in_Hc2} and of the specific heat \cite{Walmsley13}. Complementary, a behavior typically associated with strong quantum fluctuations was also observed in the normal state of P-Ba122, namely, the linear-in-$T$ resistivity \cite{AnalytisFisher2014NatPhy_QCP_BaP122,Tanatar13}. Subsequent theoretical analyses, however, \cite{Chowdhury13, Levchenko13, Hiroaki13, ChowdhurySachdevSenthil2015PRB_QCP_BaP122_Microscope_description,DKKVL-PRB15}, revealed that while a sharp enhancement of $\lambda$ is generally expected upon approaching the second-order or weakly first-order QPT from the pure SC side, due to the build up of critical AFM quantum fluctuations, a peak is not guaranteed to exist at the QPT. This is because $\lambda$ increases inside the AFM phase but does not diverge at the QPT, even if the latter is a second-order transition.
Furthermore, the detailed doping dependence of $\lambda$, including the possible peak position, was shown to depend on other non-critical properties, such as disorder and Fermi surface topology \cite{DKKVL-PRB15,Hiroaki13}.

It is therefore of general interest to establish whether the sharp peak of $\lambda$ observed in P-Ba122 is particular to this compound or a more universal property of iron-based superconductors. For instance, in other classes of unconventional superconductors, such as cuprates, a sharp peak in $\lambda$ is not observed \cite{Tallon03}. Indeed, P-Ba122 differs from most iron-pnictide materials in several key aspects. Because the substituted pnictogen atoms have little effect on the Fe plane, P-Ba122 is a very clean system, as evident from the observation of quantum oscillations \cite{Shishido10}. Moreover, P-Ba122 has a nodal superconducting gap structure, in contrast to the fully gapped superconductivity observed in iron pnictides \cite{KasaharaMatsuda2010PRB_BaP122,Tanatar10}.

Previous works on electron-doped BaFe$_2$As$_2$ did not find anomalies in $\lambda$, although the error bars were large and the step in $x$ was coarse \cite{Gordon10,Luan11,Williams10}. However, in Ba(Fe$_1$$_-$$_x$Co$_x$)$_2$As$_2$ (Co-Ba122), x-ray and neutron scattering measurements suggest a microscopic coexistence of nematic and magnetic orders with SC, supporting the existence of a QPT inside the SC phase \cite{Pratt11,Nandi10}. Other studies on optimally doped compositions report possible strong quantum fluctuations manifested in the normal state properties  \cite{Chu12,Yoshizawa12,Ning10,Tam13}, such as elastic constants \cite{Yoshizawa12}, thermopower \cite{ArsenijevicCanfield2013PRB_QCP_BaCo122_thermopower}, resistivity \cite{Taillefer2010ARCMP_Cuprate_and_others}, elasto-resistance \cite{KuoFisher2016Science_Nematic_QCP_FeSCs}, and nuclear magnetic resonance \cite{Nakai10}. In contrast to P-Ba122, Co-Ba122 displays a greater degree of disorder due to the disturbance of the Fe layers by transition metal doping, as it is clear from the large residual resistivities and low RRR \cite{Canfield10}. Moreover, in Co-Ba122 the AFM transition is split from the nematic one, whereas in P-Ba122, neutron and x-ray scattering experiments suggest that they are simultaneous and first-order \cite{Allred14}.

In this work, we use a novel minimally-invasive high sensitivity optical magnetometer based on nitrogen-vacancy (NV) centers in diamond to measure the magnitude of the London penetration depth $\lambda$ at 4.5 K across the Co-Ba122 phase diagram. The high sensitivity of the NV technique and the  precise determination of Co-doping levels via wavelength dispersive spectroscopy (WDS) allow us to clearly identify an anomalous peak in $\lambda(x)$ inside the superconducting dome near optimal doping ($x$ = 0.057). This point coincides with the extrapolated location of the AFM/SC boundary, and thus of the QPT, as determined by scattering experiments. This result demonstrates that the occurrence of a sharp peak in $\lambda$ very close to the QPT inside the dome is not limited to clean isovalently-substituted compounds with nodal superconducting gaps, but also occur in the more disordered charge-doped fully-gaped iron pnictides. This suggests that such an anomaly in the penetration depth is a more universal property of iron pnictides despite the theoretical result that a QPT alone is not enough to guarantee such an anomaly, thus shedding new light on the interplay between AFM quantum fluctuations and superconductivity in these systems.

\section*{Results and Discussion}

The lower critical field, $H_{c1}\propto \lambda^{-2}$, is obtained by detecting the onset of the first penetration of Abrikosov vortices at the sample corners as the applied magnetic field is applied to a sample cooled in zero-field to low temperature. The measurement procedure and experimental schematics of this probing scheme are discussed in detail in our previous works \cite{Joshi19,Nusran18} and summarized in \textit{Supplemental Materials}.

\begin{figure}[tb]
\centering
\includegraphics[width=8.5cm]{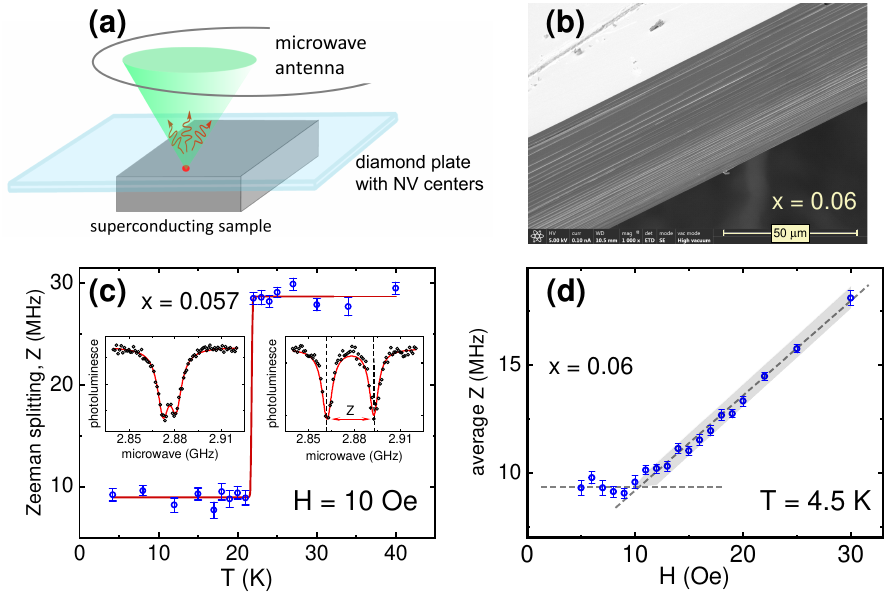}
\caption {(color online). (a) Schematics of the measurement setup. (b) SEM image of the $x=0.06$ sample showing a well-defined (001) plane and a sharp edge with the side surface corresponding to (100) plane. (c) Optically detected magnetic resonance (ODMR) splitting as a function of temperature measured on warming in a 10~Oe applied magnetic field in the $x=0.057$ sample. The two insets show the ODMR spectra below and above $T_c$. (d) Average of the smaller Zeeman splitting versus the applied magnetic field measured approximately 10 $\mu$m from the edge inside the sample with $x=0.06$.}
\label{fig1}
\end{figure}

The schematic of the experimental setup is shown in \fref{fig1}(a). The setup consists of a thin diamond plate with one surface activated with NV centers in contact with a cuboid-shaped superconducting sample. The effective demagnetization factor depends on the geometry of the sample \cite{Prozorov18}, therefore it is important to use samples with well defined shapes as determined from screening through a scanning electron microscope (SEM) \fref{fig1}(b). The magnetic induction is measured by monitoring the Zeeman splitting in optically detected magnetic resonance (ODMR). For detection of the superconducting phase transition \fref{fig1}(c), magnetic induction is measured near the center of the sample, whereas \hc measurements are performed near the sample's edge  \fref{fig1}(d). The overall measurement protocol is explained in detail in Refs.~\cite{Joshi19,Nusran18}.

\begin{figure}[tb]
\centering
\includegraphics[width=8cm]{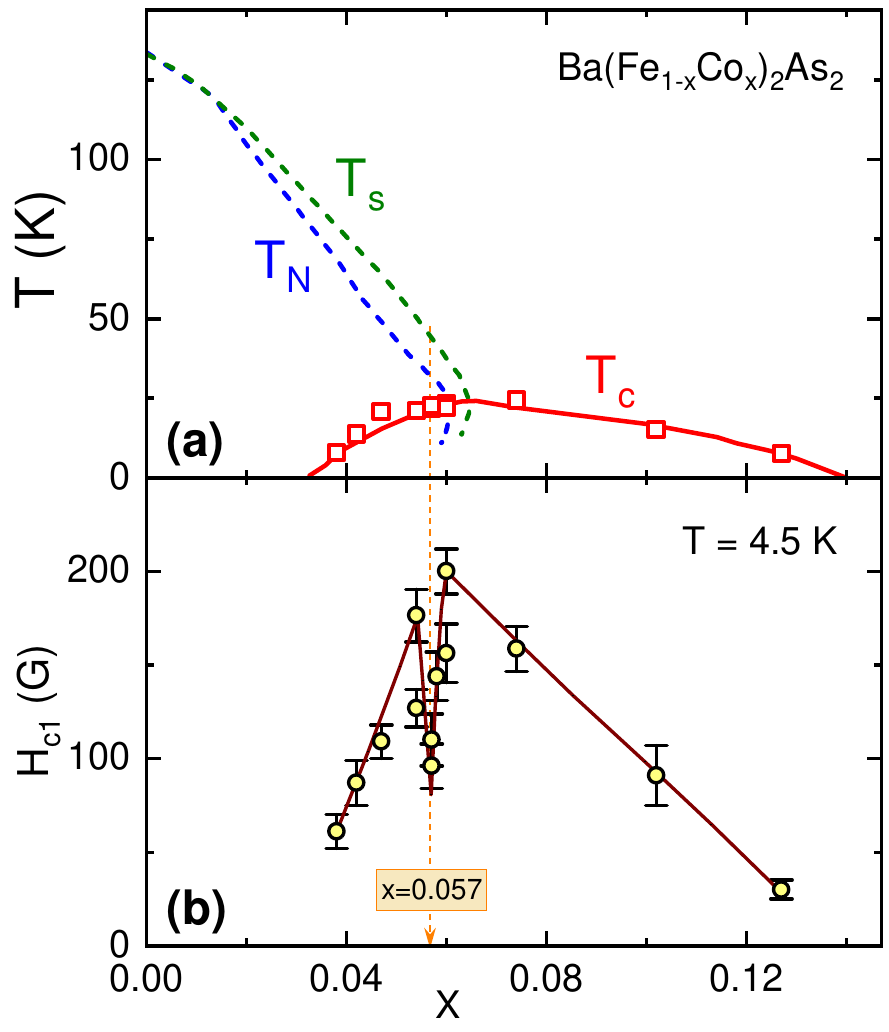}
\caption{(color online). (a) Temperature, $T$, vs. cobalt concentration, $x$, phase diagram of Ba(Fe$_{1-x}$Co$_x$)$_2$As$_2$ (Co-Ba122) from Refs.~\cite{Ni08,Nandi10, Pratt11}. Dashed lines show structural, $T_S (x)$,  and antiferromagnetic, $T_N (x)$, transition lines. Red open squares are the values of $T_c(x)$ from this work and red solid line is a guide for an eye. (b) Doping dependence of \hc at 4.5 K across the superconducting dome. A sharp dip is located at $x=0.057$.}
\label{fig2}
\end{figure}

\Fref{fig2} shows the temperature-doping phase diagram and \hc across the superconducting dome. In the underdoped region, coexisting antiferromagnetic and superconducting orders give rise to a rapid increase in \hc, as expected from general theoretical considerations \cite{Fernandes10a}. In the overdoped region, a moderate decrease in \hc is observed, likely due to increasing pairbreaking scattering with larger amount of substitutional disorder and larger superconducting gap anisotropy \cite{Tanatar10}. Most importantly, a distinct peak-shaped anomaly in an otherwise smoothly varying \hc around $x=0.057$ is clearly observed  \fref{fig2}(b). Compared with the phase diagram in \fref{fig2}(a), the anomaly is precisely at  the point where the AFM order disappears (by extrapolation to $T=0$). Measurements were performed on two different samples of the same composition, yielding consistent results.

\begin{figure}[htb]
\centering
\includegraphics[width=8cm]{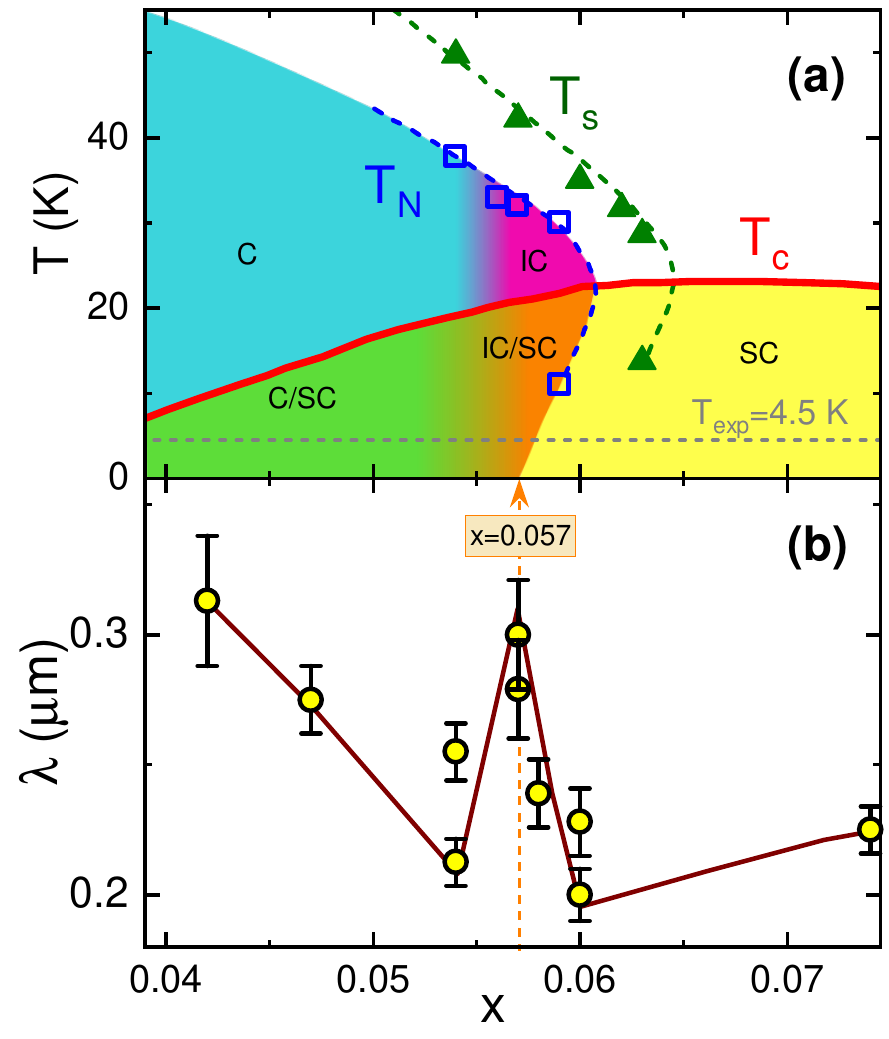}
        \caption {(color online). (a) Detailed phase diagram of Co-Ba122 in the region of structural and magnetic transition lines entering the ``dome of superconductivity". The locations of commensurate, C, and incommensurate, IC, antiferromagnetic orders are adopted from Refs.~\cite{Pratt11,Nandi10}. The horizontal dashed line shows measurement temperature of 4.5 K used in this work. (b) $\lambda$(4.5K) as a function of cobalt concentration, $x$. The peak in $\lambda$ at $x=0.057$ coincides precisely with the linear extrapolation of the back-bent $T_N(x)$ under the dome to $T=0$.}
        \label{fig3}
\end{figure}

Because our measurements are performed at a fixed $T=4.5 $ K (dashed line in \fref{fig3}(a)), the relative temperature $T/T_c(x)$ changes as function of $x$, since $T_c(x)$ varies between 24 K and 10 K in the studied range. This, however, cannot explain the observed peak, since the London penetration depth increases with the increase of $T/T_c$, and the reduced temperature is the smallest at optimal doping.
Indeed, using a crude estimate based on the two-fluid model, even at $T/T_c=0.5$ the relative change of $\Delta \lambda (0.5T_c)/\lambda(0) \leq 10\%$. Similarly, flux pinning cannot account for the peak anomaly. In our earlier work \cite{Prozorov2009c}, we showed that the critical current density, $j_c(x)$, peaks approaching optimal doping. This indicates efficient pinning on the structural domains, which become finer, so their density increases and width decreases, both leading to the enhancement of pinning. However, in any model of pinning, critical current is inversely proportional to the London penetration depth (clearly, there is no pinning if $\lambda$ diverges) and, therefore, this mechanism would result in a dip, not a peak in $j_c(x)$. Moreover, the fact that we do not see this behavior supports our assertion that we do not enter the vortex pinning regime at all and only detect the onset of flux penetration.

To pin down the position of the peak in $\lambda$, \fref{fig3}(a) zooms the temperature-doping phase diagram obtained in Refs.\cite{Nandi10,Pratt11} near the optimal doping region. The commensurate (C) antiferromagnetic order evolves into incommensurate (IC) at the very edge of the AMF phase, as expected theoretically \cite{VorontsovChubukow2010PRB_SDW_FeSCs}. $\lambda$ deduced from our \hc measurements in the corresponding region is shown in \fref{fig3}(b). An  anomalous increase in $\lambda$ is clearly visible near the composition where AFM order becomes incommensurate and eventually disappears. Note that the extrapolated position of the AFM-QPT inside the SC dome is not the same as if the extrapolation was done above the SC dome. The reason is the back-bending of the AFM transition line, which is observed by neutron scattering \cite{Nandi10,Pratt11} and attributed to the competition between AFM and SC \cite{Fernandes10}.

\begin{figure}[htb]
\centering
\includegraphics[width=8cm]{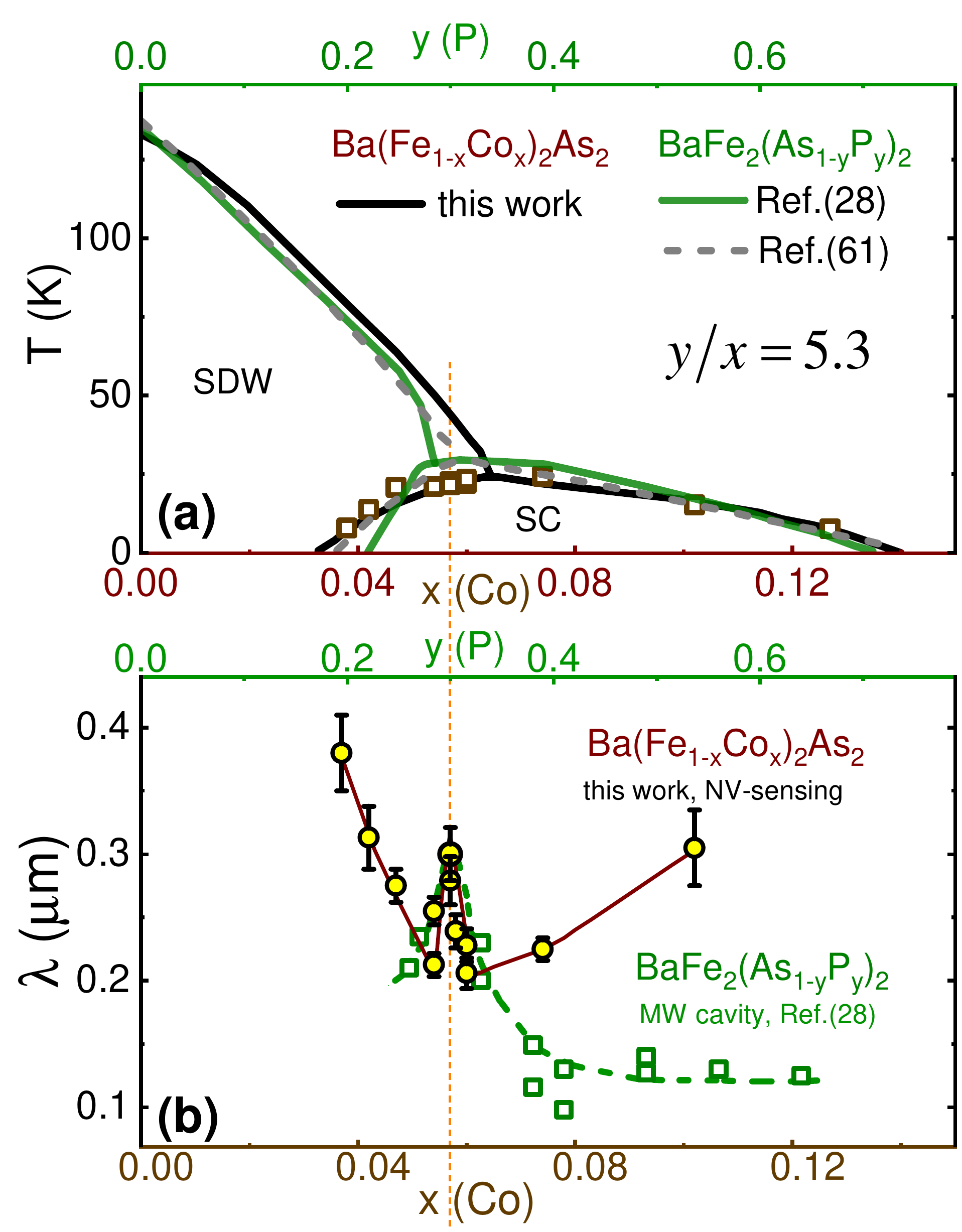}
        \caption{(color online). (a) Comparison of temperature-doping phase diagrams of Co-Ba122 (black line \cite{Nandi10}) and P-Ba122 (green solid line \cite{Hashimoto12}, and grey dashed line \cite{Hu15}). Open squares are the superconducting transition temperature, $T_c(x)$, from this work. A remarkable scaling is achieved with all lines practically coinciding without any change in the temperature axis and only phosphorus concentration, $y$, shown in the top axis, divided by the factor of 5.3. (b) Doping dependence of $\lambda$(4.5 K) across the superconducting dome measured using optical NV magnetometry (open circles, this work). For comparison, open green squares show $\lambda$(4.2 K) in P-Ba122 measured by using the microwave cavity perturbation technique.\cite{Hashimoto12}}
\label{fig4}
\end{figure}

A direct comparison of Co-Ba122 and P-Ba122 reveals striking similarities. It turns out that a simple re-scaling of the phosphorus composition (divided by a factor of 5.3) results in a good match of all principal transition lines as shown in \fref{fig4}(a), where $T_N(x)$ lines were not changed between two compounds. When plotted in the re-scaled phosphorus $y-$axis, as shown in \fref{fig4}(b), the behavior of $\lambda (x)$ near optimal doping is remarkably similar in both compounds with similar peak values of $\sim$ 300 nm. This is astounding considering how different the behavior is deep in the overdoped region. In this region, the increase of $\lambda (x)$ observed in overdoped Co-Ba122 may be attributed to a significant increase of the scattering rate due to charge doping, which suppresses the superfluid density, and to an increasing gap anisotropy. In contrast, isovalent, hence cleaner, P-Ba122 remains flat. On the underdoped side, both compositions show a steep increase of $\lambda (x)$  (for P-Ba122, see MFM measurements in Ref. \cite{Lamhot15}) due to coexisting magnetic order.

It is surprising that both compounds display such a similar behavior for the penetration depth near the putative QPT. Surprising, not only because of how different their disorder level and gap structure are (clean and nodal for P-Ba122, dirty and nodeless for Co-Ba122), but also because of the different characters of their AFM and nematic transitions. While in P-Ba122 they collapse into a single first-order transition line well before crossing the SC dome, in Co-Ba122 two separate, second-order transition lines cross the superconducting dome and continue to exist separately down to $T/T_c \approx 0.5$. For lower temperatures, the fate of these transitions is not well understood, at least experimentally, while theoretically they are predicted to merge and continue as a single weakly first-order transition line down to $T = 0$ \cite{Fernandes13}. This would imply a single QPT may exist in both P-Ba122 and Co-Ba122.

Importantly, even if a second-order QPT exists within the superconducting dome, be it AFM or nematic, theoretical analyses show that its critical fluctuations are expected to cause an enhancement of $\lambda$, without a divergence, upon approaching the QPT from the non-AFM side, but not necessarily a peak \cite{Chowdhury13, Levchenko13, Hiroaki13, ChowdhurySachdevSenthil2015PRB_QCP_BaP122_Microscope_description,DKKVL-PRB15}. This makes it even more surprising our observation that a peak in $\lambda$ exists and nearly coincides with the extrapolated AFM-QPT in two compounds as different as P-Ba122 and Co-Ba122. Interestingly, disorder has also been proposed to be an important ingredient to trigger a peak in $\lambda$, either by promoting a SC-AFM micro-emulsion with frustrated Josephson couplings between SC grains \cite{ChowdhurySachdevSenthil2015PRB_QCP_BaP122_Microscope_description}, or by tuning the balance between the competing SC and AFM orders \cite{DKKVL-PRB15}. While the residual resistivity ratios of P-Ba122 and Co-Ba122 are dramatically different, indicating that the latter is much dirtier than the former, recent nuclear magnetic resonance (NMR) measurements reported evidence of significantly inhomogeneous dynamics in both compounds \cite{Curro16}.

\section*{Conclusions}

In conclusion, the absolute value of the London penetration depth at $T=4.5$~K, $\lambda (x)$, was measured across the superconducting dome of Ba(Fe$_{1-x}$Co$_x$)$_2$As$_2$ (Co-Ba122) using sensitive minimally-invasive optical magnetometry based on nitrogen-vacancy (NV) centers in diamond. The  measurements revealed a sharp peak in $\lambda (x)$, which coincides with the quantum phase transition (QPT) found by the extrapolation to $T_N \rightarrow 0$ of the antiferromagnetic (AFM) phase boundary \emph{inside} the SC dome. This result shows that the peak in $\lambda$ is not limited to clean isovalently-substituted compounds with nodal superconducting gaps, but also exists in more disordered electron-doped compositions with fully-gaped superconductivity, suggesting that this may be a more universal and ubiquitous manifestation of a QPT in iron-based superconductors.

This puzzling observation raises important theoretical questions regarding the interplay between SC and AFM, as one does not expect that a QPT will generally lead to a peak in $\lambda (x)$ \cite{Chowdhury13, Levchenko13, Hiroaki13, ChowdhurySachdevSenthil2015PRB_QCP_BaP122_Microscope_description,DKKVL-PRB15}.
This result also suggests that the QPT inside the dome is unexpectedly robust with respect to disorder. The significance of disorder in determining the physics of QPT is usually quantified by the Harris criterion, which links the critical exponents of correlation radius, specific heat and dimensionality to satisfy a specific condition \cite{Chayes-PRL86}. However, we are unaware of its generalization to the multi-component systems and thus can not fully explore its implications to apparently universal behavior of $\lambda$. Whether the same manifestation is featured in other unconventional superconductors remains to be determined, but there is a mounting evidence of its ubiquitous nature. For example, recent NMR study of NaFe$_{1-x}$Co$_x$As, where $T_N$ and $T_S$ lines are significantly separated on a $T-x$ diagram, finds two peaks in $\lambda(x)$ at the concentrations corresponding to the extrapolation of these transitions to $T=0$ \cite{Wang18}. A peak-like feature in $\lambda (x)$ is observed by the magnetic-force microscopy in hole-doped pnictide, Ba$_{1-x}$K$_x$Fe$_2$As$_2$ (K-Ba122) \cite{Almoalem18}. On the other hand, the data available in high$-T_c$ cuprates suggest the opposite behavior, with $\lambda$ dipping, not peaking, at the putative QPT \cite{Tallon03}. Considering how few parameters can be used to probe QPT inside the superconducting state, further detailed investigations of $\lambda$ are clearly warranted to establish a full and objective picture.

\section*{Methods}

\subsection*{Sample preparation}
High quality single crystals of BaCo122 were grown by using self-flux solution growth technique as described in Ref.~\cite{Ni08}. Cobalt concentration was measured by using wavelength dispersive spectroscopy (WDS). Crystals were first cleaved with razor blade into thin plates typically 50 $\mu$m thick and two shiny cleavage surfaces corresponding to (001) plane of the tetragonal structure. Cuboid samples  with four sharp edges were further cleaved from the platelets along (100) and (010) tetragonal directions, see Fig.1(b) of the main text. Side surfaces of the cleave are of high enough quality to make optical reflectance measurements \cite{Moon13} despite notable slab structure. Quality of the edges between (001) top and (100) side surfaces was controlled by SEM imaging.  Only those samples were selected which had well-defined sharp edges as shown in Fig.1(b) of the main text and even (001) surfaces so that the sensor is in direct contact with the sample.

\subsection*{Experimental setup and determination of $\lambda$}
The nitrogen-vacancy centers are embedded in a 40~$\mu$m thick electronic-grade single crystalline diamond plate (purchased from {\it{Element-Six}}) with [100] surface. NV centers are activated only in one side at approximately 20~nm deep from the surface. This diamond plate is placed directly onto a flat surface of a superconducting sample such that the surface containing NV-centers is in direct contact with the sample. The low-temperature measurement setup is based on Attocube AFM/CFM combo. Sensor preparation, measurement protocols and experimental setup are explained in detail in Ref.~\cite{Nusran18}.

For the \hc measurements, the sample is cooled down to 4.5~K in zero magnetic field and then magnetic field is applied along the $z-$direction (crystallographic $c-$axis), perpendicular to the sample flat face. The confocal objective is focused on the NV centers at a spot right at the edge (inside) of the sample and optically-detected magnetic resonance (ODMR) splitting (proportional to the local magnetic induction) is measured. When the applied field is increased, above a field of first vortex penetration, H$_p$, Abrikosov vortices enter the sample cutting the sharp corners and the deviation of the signal from otherwise linear behavior is detected. We note that this field is different from the field of flux penetration calculated by Brandt \cite{Brandt2001} at which vortex segments meet at the center and the whole vortex is pushed into the sample interior by the Lorentz force resulting in a significant change in $M(H)$ dependence. In our case, the detected field corresponds to \hc amplified by the demagnetization correction. Specifically, using recently calculated effective demagnetization factors for $2a\times 2b\times 2c$ cuboid-shaped samples \cite{Prozorov18},

\begin{equation}\label{eq:demag}
N = \frac{1}{1+\frac{3c}{4a}(1+\frac{a}{b})},
\end{equation}

\noindent the lower critical field, \hc, and, consequently, $\lambda$, are deduced from the measured $H_p$,

\begin{equation}\label{eq:Hc1}
H_{c1} =\frac{H_p}{1-N}=\frac{\Phi_0}{4\pi\lambda^2}
\left[ \ln\left(\frac{\lambda}{\xi}\right) + 0.5 \right],
\end{equation}

\noindent where $\Phi_0$ is flux quantum and $\xi$ is coherence length and numerical factor 0.5 is from the revised calculations of \hc by C. R. Hu \cite{Hu1972}. Recently, Yip {\it et al.} used a similar approach to obtain critical fields in a single crystal of BaFe$_2$(As$_{0.59}$P$_{0.41}$)$_2$ \cite{Yip18}.

\section*{Acknowledgments}

We thank A. V. Chubukov, D. Chowdhury, M. Dzero, M. Khodas, and M. Vavilov for fruitful discussions. We thank W. E. Straszheim for performing WDS measurements.

Work in Ames was supported by the U.S. Department of Energy (DOE), Office of Science, Basic Energy Sciences, Materials Science and Engineering Division. The research was performed at Ames Laboratory, which is operated for the U.S. DOE by Iowa State University under contract DE-AC02-07CH11358. NMN acknowledges support from the Lab Directed Research and Development (LDRD) program at Ames Laboratory for the implementation of the  NV-center optical magnetometry. A.L. was supported in part by NSF CAREER Grant No. DMR-1653661 and by the U.S. Department of Energy, Office of Science, Basic Energy Sciences, under Award DE-SC0017888. R.M.F. was supported by the U.S. Department of Energy (DOE), Office of Science, Basic Energy Sciences, under award DE-SC0012336.

\section*{References}

%

\end{document}